\documentclass[prl,notitlepage,twocolumn]{revtex4}

\usepackage{amsmath}
\usepackage{amssymb}
\usepackage{bm}
\usepackage{graphicx}
\usepackage{times}

\usepackage{color}

\begin{document}

\title{High-Q supercavity modes in subwavelength dielectric resonators}

\author{Mikhail~V.~Rybin${}^{1,2}$}
\email{m.rybin@mail.ioffe.ru}
\author{Kirill~L.~Koshelev${}^{1,2}$}
\author{Zarina~F.~Sadrieva${}^{2}$}
\author{Kirill~B.~Samusev${}^{1,2}$}
\author{Andrey~A.~Bogdanov${}^{1,2}$}
\author{Mikhail~F.~Limonov${}^{1,2}$}
\author{Yuri~S.~Kivshar${}^{2,3}$}

\affiliation{$^1$Ioffe Institute, St.~Petersburg 194021, Russia}
\affiliation{$^2$ITMO University, St.Petersburg 197101, Russia}
\affiliation{$^3$Nonlinear Physics Center, Australian National University, Canberra ACT 2601, Australia}

\begin{abstract}
Recent progress in nanoscale optical physics is associated with the development of a new branch of nanophotonics exploring
strong Mie resonances in dielectric nanoparticles with high refractive index. The high-index resonant dielectric nanostructures
form building blocks for novel photonic metadevices with low losses and advanced functionalities. However, unlike extensively
studied cavities in photonic crystals, such dielectric resonators demonstrate low quality factors
($Q$-factors). Here, we uncover a novel mechanism for achieving giant $Q$-factors of subwavelength nanoscale resonators
by realizing the regime of {\em bound states in the continuum}. We reveal strong mode coupling and Fano resonances in high-index dielectric finite-length nanorods resulting in high-$Q$ factors at the nanoscale. Thus, high-index dielectric resonators represent
the simplest example of {\em nanophotonic supercavities}, expanding substantially the range of applications of all-dielectric resonant nanophotonics and meta-optics.
\end{abstract}


\maketitle


Trapping of light in localized modes is extremely important for various applications in optics and photonics including lasing~\cite{kodigala2017lasing}, sensing~\cite{gather2011single,foreman2015whispering}, harmonic generation~\cite{dominguez2011whispering,shorokhov2016miltifold}, Raman scattering~\cite{li2014single}, and photovoltaics~\cite{spinelli2012broadband,brongersma2014light}. For many optical devices, it becomes critical to localize electromagnetic fields
in small subwavelength volumes. Plasmonic structures based on metals allow subwavelength localization of light by means of surface plasmon polaritons~\cite{hess2012active}. However, metals impose inevitable losses and heating, which limit the device performance and efficiency.
In contrast, dielectric nanoparticles with high refractive index offer a novel way for the subwavelength localization of light
by employing the Mie resonances being limited only by the radiation damping~\cite{kuznetsov2016optically}. Unlike metallic nanoscale
structures, dielectric nanoparticles support both electric and magnetic Mie modes that expand substantially the applications of meta-optics~\cite{opn_2017}. Also, dielectric materials with high refractive index are available in a broad
spectral range. At the same time, the standard Mie theory predicts relatively low values of the quality factor ($Q \approx 10$) for
nanoparticles made of conventional optical materials such as Si, Ge, and AlGaAs, both in the visible and near-infrared spectral ranges.

\begin{figure}[!b]
\includegraphics{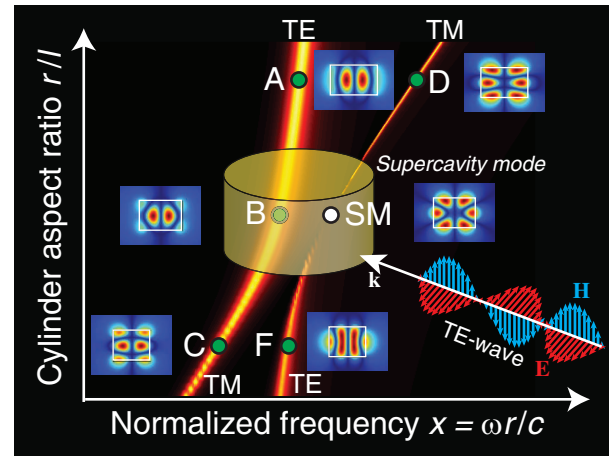}
\caption{
Illustration of strong mode coupling and bound state in the continuum supported by a high-index dielectric resonator.
Frequency of the TE (Mie-type) mode is a vertical line connecting the points A and F. When the nanorod aspect ratio varies, this line crosses with the frequency of TM (Fabry-Perot-type) mode that connects the points D and C. This avoided crossing behavior is accompanied by the formation of the supercavity mode with mixed polarization at the SM point. The curve thickness corresponds to the linewidth in the
scattering spectra. Insets show the calculated field patterns of the modes marked by filled circles.
}
\label{fig:Scheme}
\end{figure}

However, for many applications of all-dielectric nanophotonics it is very desirable to achieve higher values of the $Q$ factor. One way to enhance the $Q$ factor is to increase the size of the resonator, for example by confining waves by cavities and defects in photonic crystals~\cite{asano2017photonic} or by exploiting modes with high angular momentum known as whispering gallery modes (WGM)~\cite{righini2011whispering}. Another way is to arrange several resonators in space and excite collective modes~\cite{zywietz2015electromagnetic,shalaev2015high}. An alternative approach for enhancing the $Q$ factors is to use the so-called {\em anapole mode} with the spectrally overlapped electric and toroidal dipole modes~\cite{basharin2015dielectric,miroshnichenko2015nonradiating}. As a result, the $Q$ factor of the anapole mode realized in a dielectric resonator may exceed 30~\cite{grinblat2016efficient}. Here we suggest a novel approach based on bound states in the continuum.

The physics of bound states in the continuum (BIC) with potentially infinite lifetime attracted a lot of attentions in photonics~\cite{marinica2008bound,hsu2013observation}. Albeit true BIC can exist in structures that are infinitely extended at least at
one of spatial dimensions~\cite{hsu2016bound}, finite-size samples can support their analogue in the
form of {\em supercavity modes}~\cite{rybin2017optical}, with the $Q$ factor growing rapidly before reaching a maximum value
owing to the finite-size effects. In accord with the Fridrich-Wintgen theory of BIC~\cite{friedrich1985interfering},
the radiating tails of leaky modes can cancel out each other via destructive interference resulting in the rapid growth of
$Q$ factor. The earlier studies of high-$Q$ microcavities in the regime of avoided resonance
crossing~\cite{wiersig2006formation} can also be linked to the BIC concept~\cite{friedrich1985interfering}. In that way,
a pair of strongly coupled modes of a microcavity may interfere destructively outside the resonator \cite{wiersig2006formation}
realizing the BIC conditions. Recently, it was shown~\cite{gomis2017anisotropy} that even a simple waveguiding system can support BIC
by employing an anisotropic material and vectorial nature of electromagnetic fields. It is crucially important
that the modes with mixed polarizations may allow the substantial reduction of resonator's dimensions, and they can be
engineering to realize the supercavity regime~\cite{kivshar2017optical}.

In this Letter, we reveal that subwavelength dielectric resonators with high refractive index can support the supercavity modes
with unexpectedly high $Q$ factors. For an illustrative case of a single dielectric cylindrical resonator, this occurs when two eigenmodes with different polarizations, associated with the Mie resonances and Fabry-Perot resonances of a finite-length
nanorod, strongly interact near the avoided crossing regime (Fig. 1). Because these modes are approximately orthogonal inside
the resonator, they interfere predominantly outside, realising the so-called {\em bound states in the continuum} (BIC) and supercavity regime~\cite{rybin2017optical} when the subwavelength all-dielectric resonators demonstrate extremely high values of the $Q$ factor.

\begin{figure}[!t]
\includegraphics{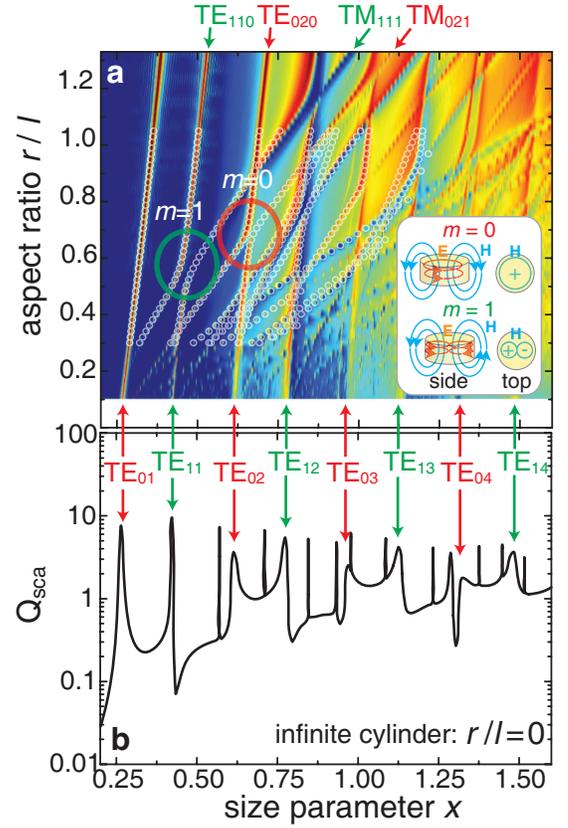}
\caption{
Dispersion of modes in a high-index dielectric resonator. (a) Plane-wave scattering spectra of a dielectric nanorod with $\varepsilon=80$ vs. the (radius-to-height) aspect ratio. Nearly vertical TE$_{mkn}$ modes correspond to the Mie resonances of an infinite nanowire [with the spectrum
shown in (b)] that can be calculated by a rigorous Mie theory. The avoided crossing of the modes with the azimuthal numbers $m=0$ and $m=1$ are marked by red and green circles, respectively. Real parts of the eigenfrequencies are obtained by the resonant-state expansion
method and are shown by small circles. Inset in (a) illustrates electric and magnetic fields for the modes with different azimuthal
numbers $m$.
}
\label{fig:ColorMap}
\end{figure}

We calculate the scattering cross-section of the dielectric cylindrical resonator as a function of its aspect ratio $r/l$. For generality, we use the normalized size parameter $x = kr$ being a product of the wavenumber $k$ and resonator radius $r$. To reveal the strong mode coupling effects, we start by setting a high value of the resonator permittivity $\varepsilon = 80$. The color map shown in Fig. 2(a) exhibits two families of resonances with the linear dependencies on the aspect ratio $r/l$. This includes the slow-varying TE-polarized (Mie-like) modes mapped to the Mie resonances of an infinitely long cylinder [see Fig.~2(b)], and rapidly-varying (Fabry-Perot-like) modes with the TM polarization. Here, we employ the standard mode classification~\cite{zhang2008electromagnetic}, so that both TE and TM modes are defined by the predominant polarization; the first index $m$ corresponds to the azimuthal number of the mode, whereas other two indices $k$ and $n$ enumerate the Mie and Fabry-Perot resonances, respectively. Due to different spectral shifts of the Mie and Fabry-Perot modes when the aspect ratio varies, the modes can intersect at certain points [as shown by red and green circles in Fig. 2(a)]. The modes with the same azimuthal index $m$ interact stronger, and they undergo coupling each other with the avoid crossing scenario at special values of the aspect ratio parameter $r/l$ with nearly complete disappearance of spectral lines (see the SM point in Fig. 1). We notice that the similar spectral feature was observed in the reflectance spectra of a photonic-crystal membrane supporting BIC~\cite{hsu2013observation}.

\begin{figure*}[!t]
\includegraphics[width=1.99\columnwidth]{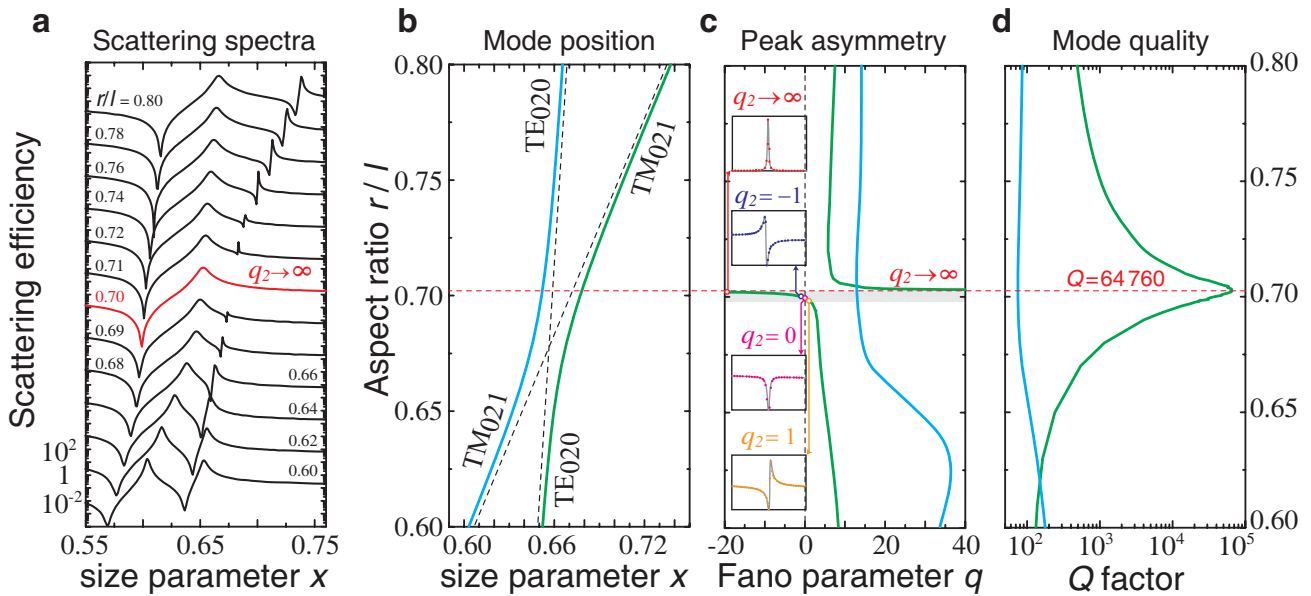}
\caption{
Characterization of the supercavity modes supported by a dielectric nanoscale resonator.
(a) Scattering spectra of a nanorod with $\varepsilon=80$ under the condition of the supercavity mode ($m=0$). Supercavity mode occurs at $r/l=0.701$ (shown in red). Spectra are relatively shifted by 100 dB.
(b) Frequencies of the Mie-type TE$_{020}$ and Fabry-Perot-type TM$_{021}$ modes in the avoided crossing regime for the modes with the azimuthal number $m=0$. (c) Fano parameter for both the modes. Insets show higher-frequency spectral features in the linear scale for the values in a narrow range of $r/l$ marked by orange, magenta, blue and red circles ($r/l=0.698$, $q=1$; $r/l=0.699$, $q=0$; $r/l=0.700$, $q=-1$; $r/l=0.703$, $q\to\infty$). (d) Calculated $Q$ factor for a typical supercavity with the maximum close to $Q=65000$.
Data in (b)-(d) are obtained by fitting to the Fano formula. Cyan and green curves correspond to the low and high frequency modes, respectively.
}
\label{fig:Spectra}
\end{figure*}

Figure~3 shows the spectra corresponding to the avoided crossing that occurs for TE$_{020}$ (Mie-like) and TM$_{021}$ (Fabry-Perot-like)
modes of the dielectric nanoresonator. In the scattering spectra [see Fig.~3(a)], each mode is manifested by an asymmetrical Fano-line shape. We study the spectra in detail through the line characteristics such as the central position, damping (proportional to the inverse value of the $Q$ factor), and Fano asymmetry parameter calculated with the help of the classical Fano formula~\cite{g020}.
Outside the avoided crossing regime, the frequency shifts of both Mie and Fabry-Perot modes are described by linear relations [see Figs.~2(a)~and~3(b). For shorter resonators, when $r/l > 0.72$, the spectrum exhibits a pair of asymmetric resonances.  The lower-frequency feature corresponds to the excitation of the TE$_{020}$ (Mie-like) mode at $x=0.66$, and the higher-frequency mode is the TM$_{021}$ (Fabry-Perot type) mode at $x=0.73$. For longer resonators, when $r/l < 0.65$,  the sequence of the modes is reversed: the Fabry-Perot mode appears at $x=0.60$, while the Mie-type mode appears at almost the same position $x=0.64$. For $0.65 < r/l <  0.72$,  the modes undergo a strong coupling accompanied by a typical avoided crossing.

\begin{figure}[!t]
\includegraphics[width=\columnwidth]{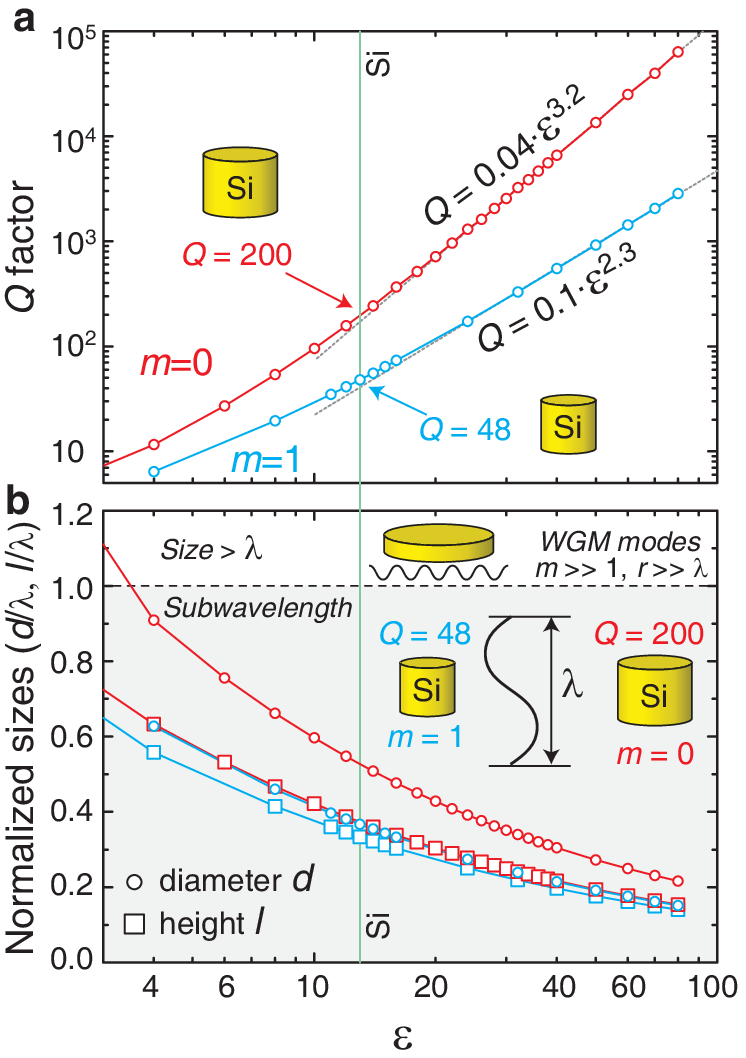}
\caption{
Calculated $Q$ factors of the supercavity modes realized in subwavelength dielectric resonators.
(a) Dependence of the $Q$ factor of the modes with the azimuthal numbers $m=0$ and $m=1$ on the dielectric permittivity.
For high values of permittivity, the $Q$ factor demonstrates a power growth.
(b) Relative dimensions of a dielectric cylindrical resonator supporting a supercavity mode. All sizes are normalized by the
resonant wavelength. A resonator becomes subwavelength below a boundary marked by a horizontal line. Diameters depicted by circles;
heights are shown by squares. Insets illustrate the cases of a silicon resonator for the $m=0$ and $m=1$ modes.
All data in (a) and (b) are calculated by the resonant-state expansion method. The case of silicon is marked by vertical lines.
}
\label{fig:Eps}
\end{figure}

Since the modes correspond to asymmetric Fano-like lineshapes, the damping parameter cannot be determined directly from the linewidth, but it can be calculated together with the Fano parameter $q$ by fitting to the Fano formula. Dependence of these two parameters on the resonator aspect ration $r/l$ is shown in Figs. 3(c,d).  For small values of the aspect ratio ($r/l < 0.62$),  both modes are characterized by positive $q$ and comparable damping parameters. When the aspect ratio decreases, the characteristics of the lower-frequency mode do not vary substantially. In contrast, for the high-frequency mode both the parameters change dramatically demonstrating the resonant features in a narrow range of the aspect ratio values, $0.69<r/l<0.71$.  At $r/l=0.703$,  the asymmetry parameter tends to infinity, and then it becomes negative. Insets in Fig.~3(c) show a change of the scattering spectra in a small range of the aspect ratio values, $0.698<r/l<0.703$. At the same time, the $Q$ factor exhibits a typical dependence of the BIC state~\cite{hsu2016bound,rybin2017optical}, with the maximum value $Q=65000$ limited by finite-size effects.

We notice that the extreme values of both damping and Fano parameters are realized at $r/l=0.703$. In the weak-coupling regime, the Fano asymmetry parameter is known to correspond to the damping rate~\cite{rybin2016purcell}, and the case $|q|\to\infty$ corresponds to uncoupling of the resonance from the continuum~\cite{riffe2011classical}.  Hence, we expect similar features also in the case of the strong coupling regime, and that suggests that the Fano parameter may be employed for the analysis of specific properties of the BIC states. Indeed, the relation between the Fano resonance and bound states in the continuum was discussed earlier in the context of electronic transport~\cite{kim1999resonant}, and the limit $q\to\infty$ corresponding to the BIC condition was termed as ``collapse of Fano resonance''. Also, the asymmetric Fano lineshape was identified  in the reflection spectra from a photonic-crystal membrane sustaining BIC~\cite{hsu2013observation}. Here, we show that an infinite value of the Fano parameter corresponds to BIC or supercavity mode.

To verify that the effective damping of the higher-frequency mode indeed vanishes, we calculate eigenfrequencies of the resonator.  We employ
a rigorous method of resonant-state expansion~\cite{doost2014resonant}. The real parts of calculated eigenfrequencies are added to Fig.~2(a). Their positions are in an excellent agreement with the features in the scattering spectra. The analysis of the imaginary parts of the resonator eigenfrequencies confirms that in the vicinity of the avoided crossing point the radiation losses of one mode decreases substantially revealing its BIC nature. In particular, for $\varepsilon=80$ the $Q$ factor of the supercavity mode is found to be 63700, being in an agreement with the results obtained from a fitting of the scattering spectra.

Finally, we discuss the conditions of the supercavity regime and BIC states in dielectric resonators. For the values of permittivity from the interval $4 \le \varepsilon \le 100$, we calculate the resonator eigenvalues by means of the resonant-state expansion method~\cite{doost2014resonant} as functions of the aspect ratio $r/l$. We focus on two anti-crossing regions marked by circles in Fig. 2(a) for the azimuthal numbers $m=0$ and $m=1$, respectively. For each value of permittivity, the $Q$ factor behaves similar to the dependence presented in Fig. 3(d). The maximum value of $Q$ (in the regime of the supercavity mode) vs. $\varepsilon$ is shown in Fig.~4(a). For $\varepsilon>10$, this dependence is well approximated by a power law, $Q=0.04\cdot\varepsilon^{3.2}$, for $m=0$, and $Q=0.1\cdot\varepsilon^{2.3}$, for $m=1$. In particular, for a Si nanoresonator with $\varepsilon=13$, the lower-frequency supercavity mode with $m=1$ demonstrates $Q$ about 50 that is at least an order of magnitude higher than the $Q$ factor of a Si sphere $Q=13.4$, for the magnetic dipole mode, and $Q=6$, for the electric dipole mode. In contrast, the supercavity mode with $m=0$ has $Q$ about 200, that is sufficient for strong nonlinear effects at the nanoscale \cite{shorokhov2016miltifold}.

Figure 4(b) demonstrates the sizes (diameter and height) of a cylindrical resonator (normalized by the wavelength) as functions of the material
permittivity. Four curves correspond to two diameters (circles) and two heights (squares) for the regime of the supercavity mode with
azimuthal numbers $m=0$ (red) and $m=1$ (cyan), respectively. Resonators operate in the subwavelength regime at moderate values of $\varepsilon>3.5$. The Si resonators are shown in the insets. Both dimensions are approximately half-wavelength, being much smaller than WGM resonators exploiting high-order $m$ modes.

The mode interference associated with the BIC state can be described in terms of the Friedrich-Wintgen theory~\cite{friedrich1985interfering}
developed earlier for electronic BIC states. This theory assumes that a pair of discrete states interact through the continuum of delocalized states. This interaction shifts the energy levels of the discrete states, and at certain conditions the states undergo a strong mode coupling. As a result, the damping rate of one state increases while the damping rate of the other state decreases until it vanishes and BIC appears. In photonic systems, a pair of resonator's modes can be described by the Friedrich-Wintgen theory provided the size of the optical modes is negligible relative to the wavelength in the media. The calculated damping rates shown in Fig. 3(d) are in an excellent agreement with the predictions of the Friedrich-Wintgen theory. When the dielectric permittivity increases, one achieves higher $Q$ factors [see Fig. 4(a)] because the modes become close to the theoretical idealization. This theory suggests how to engineer the resonator properties for achieving
really high $Q$ factors for simple geometries.

In summary, we have revealed that subwavelength High-index dielectric resonators can support the supercavity modes with high $Q$ factors, provided their parameters are engineered to match the conditions of bound states in the continuum. These results were predicted by analysing the scattering spectra of resonators and also calculating the mode eigenvalues with the resonant-state expansion method for open electrodynamic systems.
For Si resonators, our approach allows engineering supercavity modes to achieve $Q \sim 200$, that is sufficient for nonlinear optics and quantum nanophotonics. We believe that our finding will open up new horizons for active and passive nanoscale metadevices including low-threshold nanolasers, biosensors, on-chip parametric amplifiers, and nanophotonics quantum circuits.

{\em Acknowledgments.} The authors acknowledge discussions with R.W.~Boyd, C.~De Angelis, A. Fratalocchi, J.S. Totero Gongora, A.A.~Kaplyanskii, and J.~Sipe. This work has been supported  by the Ministry of Education and Science of the Russian Federation (3.1500.2017/4.6), the Russian Foundation for Basic Research (16-02-00461), Russian Science Foundation (17-12-01581), and the Australian Research Council.


\end{document}